\documentclass[%
aps,
jmp,%
amsmath,amssymb,
reprint,%
]{revtex4-1}
\usepackage{graphicx}
\usepackage{multirow}
\usepackage{diagbox}
\usepackage{subfigure}
\usepackage{dcolumn}
\usepackage{bm}
\usepackage[mathlines]{lineno}
\usepackage{epstopdf}
\usepackage{amsmath}
\usepackage[colorlinks]{hyperref}
\hypersetup{
	colorlinks = true,
	citecolor = {blue},
	linkcolor = {red}
}

\newcommand{\blk}{\color{black}}

\begin{document}
	\title{Experimental study of secure quantum key distribution with source and detection imperfections }
	\author{ Ye Chen,$^{1}$ Chunfeng Huang,$^{1}$ Zihao Chen,$^{1}$ Wenjie He,$^{1}$ Chengxian Zhang,$^{1}$ Shihai Sun$^2\dagger$ and Kejin Wei$^{1,*}$ }
	
	\address{
		$^1$Guangxi Key Laboratory for Relativistic Astrophysics, School of Physical Science and Technology,
		Guangxi University, Nanning 530004, China\\		$^2$School of Electronics and Communication Engineering, Sun Yat-Sen University, Shenzhen 518107, China\\
		$\dagger$sunshh8@mail.sysu.edu.cn\\
		$^*$kjwei@gxu.edu.cn
	}
	\date{\today}
	
	\begin{abstract}
		
		The quantum key distribution (QKD), guaranteed by the principle of quantum physics, is a promising solution for future secure information and communication technology. However, device imperfections compromise the security of real-life QKD systems, restricting the wide deployment of QKD. This study reports a decoy-state BB84 QKD experiment that considers both source and detection imperfections. In particular, we achieved a rigorous finite-key security bound over fiber links of up to 75 km by applying a systematic performance analysis. Furthermore, our study considers more device imperfections than most previous experiments, and the proposed theory can be extended to other discrete-variable QKD systems. These features constitute a crucial step toward securing QKD with imperfect practical devices.

	\end{abstract}
	
	\maketitle
	
	\section{introduction}
	Quantum key distribution (QKD) has attracted significant interest as an  information-theoretic security communication technology. With numerous effort, QKD has been widely proven in theory~\cite{1999Lo,2000Shor,2004Gottesman} and has been experimentally demonstrated in different scenarios, such as fiber-based~\cite{Liuyang2010,2012wangshuang,2018Yuan,2018Boaron2,2020Agnesi,2021zhou-MDI} and free-space channels~\cite{2017Takenaka-free,2017Liao2,2020chenhuan}. Various QKD networks have been reported worldwide~\cite{2008Peev,2010Chentengyun,2011Sasaki,2014wang,2019Dynes-network,2021Chenyuao-network}. Recently, the QKD distance has been increased to 830 km for fiber spools~\cite{2022wamg-TF} based on twin-field QKD~\cite{2018Lucamarini}.
	
	The security of QKD is guaranteed, assuming that the features of real-life devices are in line with the theoretical models in security proofs~\cite{GLLP}. However, the existing imperfections in practical components bridge a gap between theory and practice. This aspect has opened several considerable loopholes to eavesdropping by the eavesdropper (Eve). Indeed, several quantum hacking attacks exploiting such realistic security loopholes have been reported~\cite{2010Lydersen-attack,2018Qiang-hacking,2018Yoshino-hacking,2019Wei-hacking,2020Huanganqi-hacking,2020Pang-hacking}. See~\cite{2014Lo-review,2020Xu,2020Piradola-review,2022Sun-review} for a literature review involving this topic. 
	
	To recover the security of actual QKD implementations, several important approaches, such as device-independent QKD~\cite{2007Acin,2021Schwonnek-DI,2022-XU-DI}, measurement-device-independent QKD~\cite{2012Lo,2012Braun-MDI}, and security patch, were proposed. Among them, the ``security patch" method, which monitors the parameters for imperfections of the system and considers these in a detailed (GLLP-style) security analysis, has attracted increasing interest. This method usually requires to modify the hardware or post-processing process in the current system. Following this line, several studies have been conducted focusing on different imperfections in the security proof, such as detection mismatch~\cite{2007Fung,2021Zhangyanbo}, source flaws~\cite{2013Yin,2014Tamaki,2019Pereira,2020pereira}, Trojan-horse~\cite{2014Lucamarini}, pattern effects ~\cite{2018Yoshino-hacking}, polarization-dependent loss~\cite{2018Lichenyang}, and distinguishable decoy states~\cite{2018Huang-Hacking}. Experimental demonstrations of QKD using such refined security proofs have also been reported~\cite{2015Xu,2016Tangzhiyuan,2016Wangchao-source,2020Zhouxingyu-source,2022Huang}. However, in most previous studies, the flaws were considered individually using different models.
	
	To overcome this problem, Sun and Xu~\cite{2021Sun} recently proposed a systematic performance analysis that considers both source and detection imperfections in one model. With this systematic performance analysis, legitimate users can achieve considerable secure key bits over long distances by measuring the required parameters to quantify the flaws of devices in real-life QKD systems. Nevertheless, a QKD experiment that implements such an advanced theory has yet to be completed. Furthermore, in~\cite{2021Sun}, the authors only applied it to a two-decoy-state case assuming that an arbitrarily large number of signals could be used for evaluating the final key bits. However, this assumption is unrealistic because QKD systems typically run in a limited time without an arbitrarily large number of received pulses. In addition, the distilled key is highly affected by statistical fluctuations. Therefore, the practicality of this protocol remains unknown.
	
	In this study, we present an experimental demonstration of decoy-state BB84 QKD by considering most source and detection imperfections. Our demonstration exploits the refined security proof in~\cite{2021Sun}, which only requires one decoy state and considers the finite-data-size effect. Furthermore, with the refined security proof, we successfully distribute secure key bits over different fibers by considering additional device imperfections. The theoretical and experimental contributions of this study are detailed below. 
	
	Theoretically, we present a refined version of that presented in~\cite{2021Sun}. The refined version requires only one decoy-state and finite signals for the parameter estimation. Note that because implementing one decoy is easier, and the one-decoy-state method outperforms the two-decoy-state method in most of experimental settings~\cite{2018Rusca}, our analysis is crucial for integrating the security proof into a practical QKD system and completing the demonstration of the advanced theory.

	Experimentally, contrasting to previous BB84 QKD demonstrations that do not individually consider source and detection imperfections, we carefully quantify the detection efficiency mismatch and almost all source imperfections, including inaccurate state preparation, distinguishable decoy states, and Trojan horses. Moreover, we consider the secret key rate estimation together. Finally, we successfully distributed secure key bits over up to 75 km of a commercial fiber spool.  
	
	The remainder of this paper is organized as follows. In Sec.~\ref{Theory}, we review the security proof in~\cite{2021Sun} and then present the one-decoy-state scheme in Sec.~\ref{one-decoy}.  In Sec.~\ref{experiment}, we illustrate our experimental setup and show the results. Finally, we summarize the study in Sec.~\ref{conclusion}.

	\section{Protocol}\label{Theory}

	In~\cite{2021Sun}, the authors used a systematic security model that considered almost all source flaws and detector efficiency mismatches. Based on the main results in~\cite{2021Sun}, considering that a weak coherent source is used, the final key rate can be written as follows:
	
	\begin{equation} \label{eq1}
		R \geq P_{succ} \mu e^{-\mu} Y_{1}^{\mu}\left[1-H\left(\delta_{p, 1}^{z, \mu}\right)\right]-Q_{\mu} f_{EC} H\left(E_\mu\right),
	\end{equation}
	where $P_{succ}$ is the probability of Bob's successful execution of the virtual $X$-measurement, a coefficient less than 1; $\mu$ is the intensity of the signal state; $Y_1^\mu$ and $\delta_{p, 1}^{z, \mu}$ are the single-photon yield and single-photon phase error rate, respectively; $f_{EC}$ is the error correction efficiency. $Q_\mu$ and $E_\mu$ are the total gain and the bit error rate of the signal state, respectively; and $H(x)$ represents the binary information entropy function.
	
	To accurately estimate the lower bound of the key rate, the legitimate users should bound Eve's information by numerically searching the following routine:
	\begin{equation} \label{eq2}
		\begin{array}{c}
			\mathop{Max}\limits_{\rho_{E}}: \delta_{p, 1}^{z, \mu} \quad \text { and } \quad \mathop{Min}\limits_{\rho_{E}}: P_{succ } \\
			\text { Subject to }: \delta_{b, 1}^{x, \mu}, \delta_{b, 1}^{z, \mu}, P_{ij,1}^{\beta \beta^{\prime}}
		\end{array}
	\end{equation}
	where $\delta_{b, 1}^{z (x), \mu}$ denotes the single-photon bit error rate in $Z(X)$-basis; $P_{ij,1}^{\beta\beta'}, \{i,j\}\in\{0,1\}$, and $\{\beta,\beta'\}\in\{z,x\}$ represent the probability that Alice sends a single-photon quantum state associated with bit $``i"$ in $\beta$-basis and Bob obtains bit $``j"$ in $\beta'$-basis. All of the above factors can be estimated by decoy-state analysis~\cite{2005Lo,2005wangxiangbin}.
	
	The values of device imperfections are connected to key rate formulas by incorporating them into the constrictions, which satisfy the following equation:
	\begin{equation} \label{eq3}
		\begin{aligned}
			P_{ij,1}^{z z} =& \operatorname{Tr}\left\{\rho_{E}\left[\left(f_{z_{i}} \otimes Z_{i j}\right)^{+} f\left(f_{z_{i}} \otimes Z_{i j}\right)\right] \otimes F_{j}^{+} F_{j}\right\} \\
			P_{ij,1}^{x x} =& \operatorname{Tr}\left\{\rho_{E}\left[\left(f_{x_{i}} \otimes X_{i j}\right)^{+} f\left(f_{x_{i}} \otimes X_{i j}\right)\right] \otimes F_{j}^{+} F_{j}\right\} \\
			&P_{ij,1}^{xx, vir}  = \frac{1}{4} \operatorname{Tr}\left\{\rho_{E}\left(Z_{i j}^{p} \otimes C^{+} C\right)\right\},
		\end{aligned}
	\end{equation}
	where $\rho_{E}=|\xi(i)\rangle\langle\xi(i)|$ denotes the density matrix of Eve's POVM operators, $f_{z_i}(f_{x_i})$ is a diagonal matrix that denotes the fidelity of the side channel state, $Z_{ij}(X_{ij})$ is the coding accuracy of the encoded quantum state in the $Z(X)$-basis, and $F_{j}^{+} F_{j}$ is the detection efficiency matrix of the detector. C is a dummy filter for estimating the phase error rate, which can be constructed using $F_{j}^{+} F_{j}$. Based on~\cite{2021Sun}, Eq.~(\ref{eq3}) can be further derived as follows:
	\begin{equation} \label{eq4}
		\begin{aligned}
			&\delta_{b, 1}^{z, \mu} = \frac{P_{10,1}^{zz} + P_{01,1}^{zz}} {P_{00,1}^{zz}+P_{01,1}^{zz}+P_{10,1}^{zz}+P_{11,1}^{zz}}
			\\\\
			&\delta_{b, 1}^{x, \mu} = \frac{P_{10,1}^{xx} + P_{01,1}^{xx}} {P_{00,1}^{xx}+P_{01,1}^{xx}+P_{10,1}^{xx}+P_{11,1}^{xx}} 
			\\\\
			\delta_{p, 1}^{z, \mu} &= \frac{P_{10,1}^{xx,vir} + P_{01,1}^{xx,vir}} {P_{00,1}^{xx,vir}+P_{01,1}^{xx,vir}+P_{10,1}^{xx,vir}+P_{11,1}^{xx,vir}} 
			\\\\
			P_{succ} &= \frac{P_{00,1}^{xx, vir}+P_{01,1}^{xx, vir}+P_{10,1}^{xx, vir}+P_{11,1}^{xx, vir}}
			{P_{00,1}^{zz}+P_{01,1}^{zz}+P_{10,1}^{zz}+P_{11,1}^{zz}}.
		\end{aligned}
	\end{equation}
	
	Here the superscript $vir$ in $P_{ij,1}^{xx,vir}$ represents the virtual $X$-basis. Hence, legitimate users can first measure the required parameters in a real-life QKD system and numerically find the required parameters for evaluating the final key rate by performing the routine of Eq.~(\ref{eq2}), under the constraints given in Eqs.~(\ref{eq3}) and~(\ref{eq4}). Finally, the final key rate can be evaluated by inputting the obtained parameters into Eq.~(\ref{eq1}). The model of imperfections from the measured parameters can be found in Appendix~\ref{AppendixA}. We remark that the framework of this security proof has a similar routine of recent well-established numerical approach~\cite{2016coles-Numerical,2018Winick-numberical}.

	\section{One-decoy method and finite-size analysis}\label{one-decoy}
	In~\cite{2021Sun}, the parameters in Eq.~(\ref{eq1}) were
	estimated using the two-decoy-state method without considering the finite size effect. Hence, this approach has a relatively complex implementation and requires infinite data for parameter estimation, challenging to perform experimentally.
	Here, we present a refined version requiring only one decoy state~\cite{2018Rusca} and finite signals for parameter estimations. 
	The proposed method significantly reduces the experimental complexity. Based on the method presented in~\cite{2018Rusca}, the final secure key is
	expressed as
	\begin{equation} \label{eq5}
		\begin{aligned}
			l \geq s_{z, 0}^{L} + &P_{succ}s_{z,1}^{L}\left(1-h\left(\delta_{p,1}^{z,\mu}\right)\right) - \lambda_{EC}\\
			&- 6\log _{2}\frac{19}{\varepsilon_{sec }} - \log_{2}\frac{2}{\varepsilon_{cor }},
		\end{aligned}
	\end{equation}
	where $s_{z,0}^L$ is the lower bound of the detection counts when Alice sends a vacuum state in the $Z$-basis and is detected by Bob, $s_{z,1}^L$ is the lower bound of the single-photon detection counts in the $Z$-basis, $\lambda_{EC}$ is the number of bits consumed in error correction, and $\varepsilon_{sec}=10^{-9}$ and $\varepsilon_{cor}=10^{-15}$ are the secrecy and correctness factors, respectively.

	When considering the finite-data effect and the distinguishable decoy states based on the analysis presented in~\cite{2021Sun,2018Rusca}, $s_{z,0}^L$ and $s_{z,1}^L$ can be estimated using the following expression:
	\begin{equation} \label{eq9}
		\begin{aligned}
			& s_{z, 0}^{L} = 
			\\
			&\max \left\{ \frac{1}{\mu-\nu} \left[\mu n_{z,\nu}^- - \nu n_{z,\mu}^+ - 2N_{zz}D_{\mu \nu} \mu\left(e^{\nu}-1\right)\right] , \right. 
			\\
			&\left.~~~~~~~~\frac{1}{\mu-\nu} \left[ \mu n_{z,\nu}^- - \nu n_{z,\mu}^+ - 2N_{zz}D_{\mu \nu} \nu \left( e^{\mu}-1 \right) \right] , 0 \right\}\\
			& s_{z, 1}^{L}  = \frac{\tau_{1}\mu}{\nu(\mu-\nu)} \left( n_{z,\nu}^{-}-\frac{\nu^{2}}{\mu^{2}} n_{z, \mu}^{+}-~~~~~ \right.
			\\
			&\left.\frac{\left(\mu^{2}-\nu^{2}\right)}{\mu^{2}} \frac{s_{z, 0}^{U}}{\tau_{0}}-2 N_{zz} D_{\mu \nu}\left(e^{\nu}-e^{\nu\left(1-\eta_{\mathrm{Bob}}^{cal} \right)}\right)\right), 
		\end{aligned} 
	\end{equation}
	where $\eta_{\mathrm{Bob}}^{\mathrm{cal}}=\eta_{\mathrm{Bob}} \times \eta_{\mathrm{d}}$. In addition, $\eta_{Bob}$ is the total transmittance of Bob, and $\eta_{d}$ is detection efficiency. $N_{zz}$ is the total number of pulses that Alice sends to the quantum state in the $Z$-basis and Bob successfully detects in the $Z$-basis. $\tau_{i}$ is the probability of sending $i$ photons, and $\mu$ and $\nu$ represent the intensity of signal state and decoy state respectively. Finally, $n_{z,k}^\pm$ is the finite-key correction of the counts in the $Z$-basis with respect to the intensity $k\in\{\mu,\nu\}$, estimated using Hoeffding's inequality as follows:
	\begin{equation} \label{eq7}
		n_{z,k}^\pm=\frac{e^{k}}{p_{k}}\left(n_{z, k} \pm \sqrt{\frac{n_{z}}{2} \log \frac{1}{\varepsilon_{1}}}\right),
	\end{equation}
	where	$p_{k}$ is the probability choice for intensity $k$, $n_{z}$ is the total count in the $Z$-basis, and $s_{z, 0}^{U}$ is the upper bound on vacuum events, which can be evaluated by
	\begin{equation} \label{eq8}
		s_{z, 0}^{U}  = 2\left(\frac{e^{k}}{p_{k}} m_{z, k}^+ + \sqrt{\frac{n_{z}}{2} \log \frac{1}{\varepsilon_{1}}}\right) ,
	\end{equation}
	where the estimate of $m_{z,k}^+$ is similar to Eq.~\eqref{eq7}, which can be determined as follows:
	
	\begin{equation}
		m_{z,k}^\pm=\frac{e^{k}}{p_{k}}\left(m_{z, k} \pm \sqrt{\frac{m_{z}}{2} \log \frac{1}{\varepsilon_{2}}}\right) .
	\end{equation}

	\begin{figure*}
		\centering
		\resizebox{\linewidth}{!}{
			\includegraphics[width=1\textwidth]{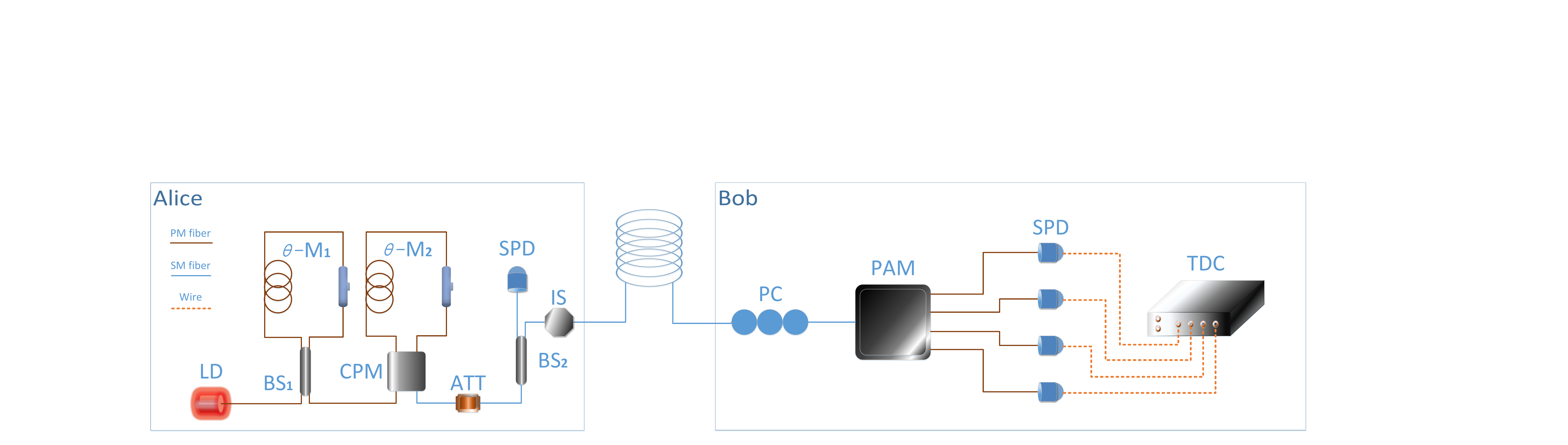} 
		}
		\caption{Schematic diagram of the BB84 QKD experimental setup. LD:  commercial laser diode; BS$_1$: beam splitter of 75:25; BS$_2$ beam splitter of 50:50; $\theta$-M$_i$: phase modulator; CPM: customized polarization module; ATT: optical attenuator; IS: optical isolation; PC: polarization controller; SPD: single-photon avalanche detector; PAM: polarization analysis module; TDC: time-to-digital converter. Polarization-maintaining fiber is used at both receiving and sending terminals.  
		}\label{fig1}
	\end{figure*}
	
	\begin{table*}[ht]
		\centering
		\caption{Required parameters for the considered imperfections, $\delta_{\beta_i}$is the modulation error in the source, $D_{\mu\nu}$ is the distance between the signal and decoy states, $\mu_{out}$ is the intensity of reflected Trojan-horse photon, and $F_{j}^{+} F_{j}, j\in\{0,1\}$ is the detection efficiency matrices of detector for bit ``$j$". In an ideal QKD system, the values of the required parameters are equal to zero in the first three imperfections, and the detection efficiency matrices are an identity matrix.  }
		\resizebox{0.7\linewidth}{!}{
			\begin{tabular}{ccc}
				\hline\hline
				Considered imperfection &  & Flaws parameters       
				\\ \hline 
				Inaccuracy of the encoded state &  & 
				$\delta_{z_0}=0,~ \delta_{z_1}=0.0726,~ \delta_{x_0}=0.0891,~ \delta_{x_1}=0.0285$ 
				
				\\
				Distinguishable decoy states &  & 
				$D_{\mu\nu} = 1.6 \times 10^{-3}$ 
				
				\\
				Trojan-horse &  & 
				$\mu_{out} = 3.6 \times 10^{-10}$ 
				
				\\
				Detection mismatch &  &
				$F_0^+ F_0 = diag[~1~,~0.6621~],~~ F_1^+ F_1 = diag[~0.3354~,~1~]$   
				
				\\ \hline\hline
			\end{tabular}
		}
		\label{table5}
	\end{table*}
	
	Here, $\varepsilon_i$ denotes the failure probability of the finite-size analysis, $i\in\{1,2\}$. To numerically search $P_{succ}$ and $\delta_{p,1}^{z,\mu}$, we need to perform decoy analysis to obtain the constraints of Eq.~\eqref{eq4}. When considering the finite data effect, Eq.~\eqref{eq4} can be rewritten as follows:
	\begin{equation} \label{eq13}
		\delta_{b, 1}^{z, \mu} \leq \frac{T_{z_0,1}^U+T_{z_1,1}^U} {s_{z_1,1}^{L}+s_{z_0,1}^{L}},
		\delta_{b, 1}^{x, \mu} \leq \frac{T_{x_0,1}^U+T_{x_1,1}^U} {s_{x_1,1}^{L}+s_{x_0,1}^{L}},
	\end{equation}
	Here, we used the following relation:
	\begin{equation} \label{eq14}
		\begin{array}{c}
			T_{z(x)_0,1}=N_{zz(xx)}P_{01,1}^{zz(xx)} 
			\\\\
			T_{z(x)_1,1}=N_{zz(xx)}P_{10,1}^{zz(xx)}
			\\\\
			s_{z(x)_0,1}=N_{zz(xx)}\left[P_{00,1}^{zz(xx)}+P_{01,1}^{zz(xx)}\right]
			\\\\
			s_{z(x)_1,1}=N_{zz(xx)}\left[P_{10,1}^{zz(xx)}+P_{11,1}^{zz(xx)}\right] ,
		\end{array}
	\end{equation}
	where $s_{\beta_{i},1}$ denote the counts that Alice sends a single-photon quantum state associated with bit $``i"$ in $\beta$-basis; $s_{\beta_{i},1}$ can be estimated using one-decoy state method using following equations~\cite{2022Huang}:
	\begin{equation} \label{eq10}
		\begin{aligned}
			s_{\beta_i,1} \geq s_{\beta_i, 1}^L  = \frac{\tau_{1} \mu}{\nu(\mu-\nu)} \left( n_{\beta_i, \nu}^{-}-\frac{\nu^{2}}{\mu^{2}} n_{\beta_i, \mu}^{+}-\right.~~~~~& \\
			\left.\frac{\left(\mu^{2}-\nu^{2}\right)}{\mu^{2}} \frac{s_{\beta_i, 0}^U}{\tau_{0}}-2 N_{\beta_i\beta'} D_{\mu \nu}\left(e^{\nu}-e^{\nu\left(1-\eta_{\mathrm{Bob}}^{cal}\right)}\right)\right) ,
		\end{aligned}
	\end{equation}
	where $n_{\beta_{i},k}^\pm$ denotes the upper and lower bound of the counts $n_{\beta_{i},k}$, $N_{\beta_i\beta'}$ represents the counts when the total number of pulses that Alice sends the quantum state in $\beta$-basis with bit $``i"$ and Bob detects in the $\beta'$-basis. Moreover, $n_{\beta_{i},k}$ can be directly obtained experimentally, $T_{\beta_{i},1}^U $ is the upper bound of single-photon error counts $T_{\beta_{i},1} $ given that Alice sends a single-photon quantum state associated with bit $``i"$ in $\beta$-basis, which can be estimated as follows:
	\begin{equation} \label{eq11}
		\begin{aligned}
			&T_{\beta_i,1} \leq T_{\beta_i,1}^U=min \{ K_m^\mu, K_m^\nu, K_m^{\mu\nu} \},
			\\\\
			&K_m^\mu= \frac{ \tau_{1}(m_{\beta_i,\mu}^{+} - e_{0}s_{\beta_i, 0}^{L}) } {\mu},
			\\
			&K_m^\nu= \frac{ \tau_{1}(m_{\beta_i,\nu}^{+} - e_{0}s_{\beta_i, 0}^{L}+2N_{\beta_i\beta'}\nu D_{\mu\nu}\eta_{\mathrm{Bob}}^{c a l}) } {\nu},
			\\
			&K_m^{\mu\nu}= \frac{ \tau_{1} \left[m_{\beta_i,\mu}^{+} - m_{\beta_i,\nu}^{-} + 2N_{\beta_i\beta'} D_{\mu\nu}\left(e^{\nu}-e^{\nu\left(1-\eta_{\mathrm{Bob}}^{\mathrm{cal}}\right)}\right) \right] } {\mu-\nu}.
		\end{aligned}
	\end{equation}
	\section{Experiment}\label{experiment}
	We implemented the above protocol using a custom-made polarization-encoding BB84 QKD system~\cite{2021Ma}. The system was realized using commercially available components with an up-gradation of security and stability. We note that our method is general and can be applied to other BB84 QKD systems. Here, we implement the developed homemade system through a case study. 
	\subsection{Setup}\label{Setup}
	A schematic of the experimental setup is shown in Fig.~\ref{fig1}. Alice used commercial lasers (LD, WT-LD200-DL, Qasky Co. LTD) to generate light pulses with a frequency of 50 MHz. The emitted light pulses were first fed into a Sagnac-based intensity modulation (Sagnac IM) \blk to actively generate two intensities for the one-decoy method. Subsequently, the light pulse passed through Saganc-based polarization modulation (Sagnac PM)\blk. A pre-calibration voltage was applied to a phase modulator using an arbitrary waveform generator (AWG) to produce standard BB84 polarization states. After that, the modulated light pulses were attenuated to the single-photon level using an attenuator (ATT).  A cascaded optical isolator (IS), with an isolation of 180.3 dB, was used in the output of Alice's station to prevent a Trojan-horse attack.
	
	At the receiving station Bob, a polarization controller (PC) was used to actively compensate for the polarization drift during transmission over the fiber channels. The received quantum states were decoded using a polarization decoding module (PAM) and detected by four infrared InGaAs single-photon detectors (SPDs, WT-SPD2000, Qasky Co. LTD). The average efficiency of SPDs was approximately $8.5\%$, and the dark count rate was $10^{-6}$. Electrical signals from the detector were recorded using a time digital converter (TDC, quTDC100, GmbH), which was then processed by a computer. After careful calibration of the system, an optical misalignment error rate of $1\%$ was obtained.

	\subsection{Implementation}\label{Implementation}

	We first experimentally quantified the required parameters for the considered imperfections in the proposed security proof, including the inaccuracy of the encoded quantum state, distinguishable decoy states, Trojan-horse, and detection efficiency mismatch. The measurement results are summarized in Table~\ref{table5}, and the detailed measurement for each imperfection and the incorporation of such parameters into the security proof can be found in Appendix ~\ref{AppendixA}.

	\begin{table*}[]
		\centering
		\caption{Implemented parameters and experimental results. $Loss$ is the channel loss, $N$ denotes the total number of pulses, $\mu$ and $v$ are the intensity of the signal state and the decoy state, respectively. $P_\mu$ is the probability of the signal state, $P_v$ is the probability of the decoy state, $E_\mu$ is the bit error rate of the signal state, and $l$ is the secret key rate.}
		\resizebox{\linewidth}{!}{
			\begin{tabular}{ccccccccccccc}
				\hline\hline
				\multicolumn{2}{c}{Channel} &~~~~~&      \multicolumn{5}{c}{Parameters}       &~~~~~& \multicolumn{4}{c}{Results}
				\\ \cline{1-2} \cline{4-8} \cline{10-13}
				~$L(km)$ ~&~ $Loss(dB)$        ~&     &    $N$ & $\mu$ & $v$ & $P_\mu$ & $P_v$   &     & $P_{succ}$ & $\delta_{p,1}^{z,\mu}$ & $E_\mu$ & $l$       
				\\ \hline 
				25 & 4.736 &&~ $1.00653\times10^{10}$ ~&~ 0.45 ~&~ 0.1125 ~&~ 0.84 ~&~ 0.16 ~&&~ 0.5360 ~&~ 0.0243 ~&~ $1.11\%$ ~&~ $8.738\times10^4$~ 
				\\
				50 & 9.602 &&~ $1.00365\times10^{10}$ ~&~ 0.43 ~&~ 0.1075 ~&~ 0.76 ~&~ 0.24 ~&&~ 0.5946 ~&~ 0.0525 ~&~ $1.03\%$ ~&~ $1.955\times10^4$~
				\\
				75 & 15.08 &&~ $1.00481\times10^{10}$ ~&~ 0.40 ~&~ 0.1000 ~&~ 0.54 ~&~ 0.46 ~&&~ 0.5601 ~&~ 0.1745 ~&~ $1.34\%$ ~&~ $7.729\times10^2$~ 
				\\ \hline\hline
			\end{tabular}
			\label{table1}
		}
	\end{table*}
	
	We successfully implemented our protocol in the finite-key regime over a commercial fiber at 25 km, 50 km, and 75 km. We optimized the implementation parameters by considering the measurement results of the evaluated imperfections. Moreover, we obtained the optimal implementation parameters, including the intensities of the signal and decoy states and the probabilities of sending them for each distance. 
	For example, our optimized parameters for 75 km were $\mu = 0.40$, $\nu=0.10$, and $P_\mu = 0.84$. Further details of the implemented parameters for each distance are listed in Table ~\ref{table1}.

\begin{figure}
		\centering
		\includegraphics[width=\linewidth]{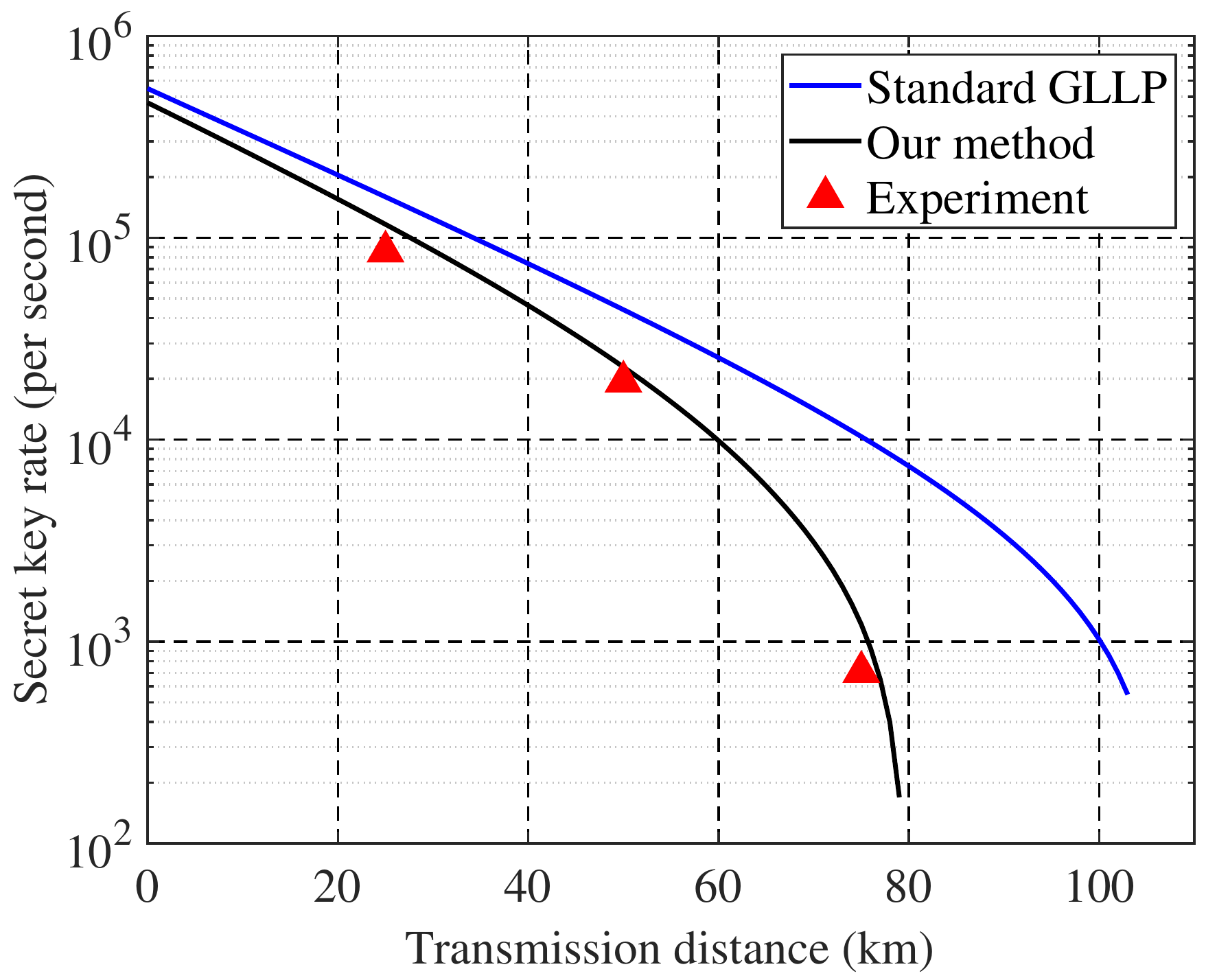}
		\caption{Experimental secret key rates (red triangle) over fiber lengths of 25, 50, and 75 km. The  black (lower) and blue (upper) lines \blk represent the simulated secret key rate using the proposed method and the standard GLLP analysis without considering imperfections. }
		\label{fig2}
	\end{figure}

	For each distance, a total of $N=10^{10}$ was sent. We collected all counts and post-processed the data for the secret key rate estimations. Detailed raw data are provided in Appendix~\ref{AppendixB}. The obtained quantum bit error rate of our system was below $1.5\%$. Substituting the experimental data into the one-decoy method presented in Sec.~\ref{one-decoy} and performing the numerical search routine of Eq.~(\ref{eq2}), we got the experimental results listed in Tab.~\ref{table1}. The results are shown in Fig.~\ref{fig2}. Based on the proposed security proof, we can distill a secret key rate  of 773 bps at a distance of up to 75 km. The security of these keys considered almost all source flaws and detection mismatch.

	In addition, we obtained the simulation results using standard GLLP analysis without considering any imperfections for comparison purposes. The simulation exploited the experimental parameters of the proposed setup, and the parameters for the decoy-state method were optimized. As shown in Fig.~\ref{fig2}, because the proposed approach considers more imperfections, our method has an acceptable penalty for the obtained secret key rate and the achievable distance.

	\section{Conclusion}\label{conclusion}

	We developed a decoy-state BB84 QKD experiment that considered almost all source flaws and detection mismatches. The study relied on the security proof reported in~\cite{2021Sun} with a simpler decoy method and considering the finite-size effect.  Finally, we successfully generated secure key bits of 773 bps in fiber links up to 75 km\blk. Compared with previous experiments that only considered a single flaw, our results proved the feasibility of distributing secure key bits over a long distance, jointly considering source and detection imperfections.

	Our method, including parameter estimations, finite-key analysis, the quantification of device imperfections, and the implement, is general. Hence, it is a interesting  future research that applies the proposed method to other QKD systems, such as a chip-based system~\cite{2017Sibson-chip,2016Ma}, to improve the security of key bits. Furthermore, considering the obtained key rate has a certain reduction compared with GLLP analysis without considering imperfections. It would be interesting to improve the secret key rate by introducing state-of-art techniques. For example, in our work, we exploited one-decoy method because of easy implementation.  introducing two-decoy method, which has been proved to provide a tighter bound than one-decoy method, would an efficient way;  Ref.~\cite{2014Tamaki,Navarrete_2022} provides a loss tolerant protocol which has a  better performance than the traditional GLLP analysis with leaky source. It is possible to use the loss tolerant protocol to make the key rate performance better.

	\section{Acknowledgments}
	This study was supported by the National Natural Science Foundation of China (Nos. 62171144,  62031024, and 62171485), the Guangxi Science Foundation (No.
	2021GXNSFAA220011), and the Open Fund of IPOC (BUPT) (No. IPOC2021A02).

	\appendix

	\section{The details of quantifying the considered imperfections}\label{AppendixA}

	In this appendix, we present the proposed method for measuring the required parameters to quantify the considered imperfections. In addition, we describe how to construct the density matrix used in the security proof.

	\subsection{Inaccuracy of encoded state}\label{Inaccuracy of encoded state}

	We quantified the inaccuracy of the encoded state by measuring the modulation error $\delta_{\beta_{i}}$ in the  Sagnac PM\blk, as shown in Fig.~\ref{fig1}. $\theta_{\beta_{i}}$ is defined as the difference between the applied and ideal phases $\theta\in\{0,\pi/2,\pi,3\pi/2\}$ when preparing the state $\beta_{i},\beta\in\{z,x\},i\in\{0,1\}$. In particular, the process is as follows. First, we determine the optimal voltages for modulating the expected polarizations following a custom calibration procedure. Subsequently, Alice connects directly to Bob via an attenuator. Subsequently, Alice scans the optimal voltages to her Sagnac-based polarization modulator and records the detection counts of the two detectors in the $Z$-basis measurement. These counts are represented by $D_{1,\theta}$ and $D_{2,\theta}$. The measurement results are listed in Tab.~\ref{table3}.

	\begin{table}[]
		\centering
		\caption{Raw data and modulation errors for state preparation. The detection efficiencies of detector 1 (2) are $8.25 \%$ and $8.76 \%$, respectively.}
		\begin{tabular}{@{} ccccc @{}}
			\hline \hline
			Polarization        &~~~$\theta$~~~  & ~~$D_{1, \theta}$~~ & ~~$D_{2, \theta}$~~ & $\delta_{\beta_i}^U$ \\ 
			\hline 
			$H(z_0)$            &0         & 824       & 348092    & /        \\ 
			$P(x_0)$            &$\pi/2$   & 146280    & 207533    & 0.0726   \\ 
			$V(z_1)$            &$\pi$     & 285143    & 3244      & 0.0891   \\ 
			$M(x_1)$            &$3\pi/2$  & 159280    & 189432    & 0.0285   \\ 
			\hline \hline
		\end{tabular}
		\label{table3}
	\end{table}

	The upper bound of $\delta_{\beta_{i}}$ is given as follows:
	\begin{equation} \label{eqa1}
		\delta_{\beta_i} \leqslant \delta_{\beta_{i}}^U = \frac{1}{2}
		\left|
		\theta -2 \arctan \left(
		\sqrt{\frac
			{\left(D_{1, \theta}^+ - D_{1,0}^- \right) / \eta_{d1} }
			{\left(D_{2, \theta}^- - D_{1,0}^+ \right) / \eta_{d2} }
		},
		\right)
		\right|
	\end{equation}
	where $D_{i, \theta}^\pm = D_{i, \theta} \pm \sqrt{D_{i, \theta} / 2 \ln (1 / \varepsilon)}$ denotes the lower and upper bounds of $D_{i,\theta}$, which is bounded by Hoeffding's inequality, and $\eta_{d_1}=8.25\%$ and $\eta_{d_2}=8.76\%$ are the detection efficiencies of the two detectors. We set $\varepsilon=10^{-10}$.
	
	To construct the density matrix $Z_{ij}$ and $X_{ij}$  used in Eq.~(\ref{eq3}) according to~\cite{2021Sun}, the state-preparation flaw parameter is defined as follows:
	\begin{equation}\label{varepsilon}
		\varepsilon_{\beta_{i}} = \sin ^{2}\left( \delta_{\beta_{i}} \right) / \eta_{c},
	\end{equation}
	where $\eta_{c}$ denotes channel loss. Subsequently, by inputting $\delta_{\beta_{i}}$ into Eq.~(\ref{varepsilon}), we obtain the coding accuracy matrices $Z_{ij}$, expressed as follows:
	\begin{equation} \label{eqa3}
		\begin{aligned}
			Z_{00}& = \left[\begin{array}{llll}		\sqrt{1-\varepsilon_{z_{0}}} & \sqrt{\varepsilon_{z_{0}}} & 0 & 0
			\end{array}\right] 
			\\	&= \left[\begin{array}{llll}	1 & 0 & 0 & 0
			\end{array}\right]
			\\\\	Z_{01}& = \left[\begin{array}{llll}0 & 0 & \sqrt{1-\varepsilon_{z_{0}}} & \sqrt{\varepsilon_{z_{0}}}
			\end{array}\right] 
			\\
			&= \left[\begin{array}{llll}	0 & 0 & 1 & 0
			\end{array}\right]
			\\\\
			Z_{10}& = \left[\begin{array}{llll}
				\sqrt{\varepsilon_{z_{1}}} & \sqrt{1-\varepsilon_{z_{1}}} & 0 & 0
			\end{array}\right] 
			\\
			&= \left[\begin{array}{llll}
				\sqrt{7.9 \times 10^{-3}/\eta_{c}} & \sqrt{1-7.9 \times 10^{-3}/\eta_{c}} & 0 & 0
			\end{array}\right]
			\\\\
			Z_{11}& = \left[\begin{array}{llll}
				0 & 0 & \sqrt{\varepsilon_{z_{1}}} & \sqrt{1-\varepsilon_{z_{1}}}
			\end{array}\right] 
			\\
			&= \left[\begin{array}{llll}
				0 & 0 & \sqrt{7.9 \times 10^{-3}/\eta_{c}} & \sqrt{1-7.9 \times 10^{-3}/\eta_{c}}
			\end{array}\right]
			.
		\end{aligned}
	\end{equation}
	The density matrix of the $X$-basis can be obtained using the transpose matrix $U$ similarly to $Z_{ij}$, which is expressed as follows:
	\begin{equation} \label{eqaXij}
		\begin{aligned}
			X_{00}& = \left[\begin{array}{llll}
				\sqrt{1-\varepsilon_{x_{0}}} & \sqrt{\varepsilon_{x_{0}}} & 0 & 0
			\end{array}\right] \times U
			\\
			&= \left[\begin{array}{llll}
				\sqrt{1-5.3 \times 10^{-3}/\eta_{c}} & \sqrt{5.3 \times 10^{-3}/\eta_{c}} & 0 & 0
			\end{array}\right] \times U
			\\\\
			X_{01}& = \left[\begin{array}{llll}
				0 & 0 & \sqrt{1-\varepsilon_{x_{0}}} & \sqrt{\varepsilon_{x_{0}}}
			\end{array}\right] \times U
			\\
			&= \left[\begin{array}{llll}
				0 & 0 & \sqrt{1-5.3 \times 10^{-3}/\eta_{c}} & \sqrt{5.3 \times 10^{-3}/\eta_{c}}
			\end{array}\right] \times U
			\\\\
			X_{10}& = \left[\begin{array}{llll}
				\sqrt{\varepsilon_{x_{1}}} & \sqrt{1-\varepsilon_{x_{1}}} & 0 & 0
			\end{array}\right] \times U
			\\
			&= \left[\begin{array}{llll}
				\sqrt{8.1 \times 10^{-4}/\eta_{c}} & \sqrt{1-8.1 \times 10^{-4}/\eta_{c}} & 0 & 0
			\end{array}\right] \times U
			\\\\
			X_{11}& = \left[\begin{array}{llll}
				0 & 0 & \sqrt{\varepsilon_{x_{1}}} & \sqrt{1-\varepsilon_{x_{1}}}
			\end{array}\right] \times U
			\\
			&= \left[\begin{array}{llll}
				0 & 0 & \sqrt{8.1 \times 10^{-4}/\eta_{c}} & \sqrt{1-8.1 \times 10^{-4}/\eta_{c}}
			\end{array}\right] \times U
			,
		\end{aligned}
	\end{equation}
	and	
	\begin{equation} \label{eqa4}
		\begin{aligned}
			U& = \frac{1}{2}\left[\begin{array}{rrrr}
				1 &  1 &  1 &  1  \\
				1 & -1 &  1 & -1  \\
				1 &  1 & -1 & -1  \\
				1 & -1 & -1 &  1
			\end{array}\right] .
		\end{aligned}
	\end{equation}
	
	Note that similar methods have been used to quantify the modulation errors in a BB84 phase-encoding system~\cite{2015Xu}, a polarization-encoding measurement-device-independent system~\cite{2016Tangzhiyuan}, and the polarization-dependent loss in a BB84 polarization-encoding system~\cite{2022Huang}.

	\subsection{Distinguishable decoy state}\label{Distinguishable decoy state}
	
	\begin{figure}
		\centering
		\includegraphics[width=\linewidth]{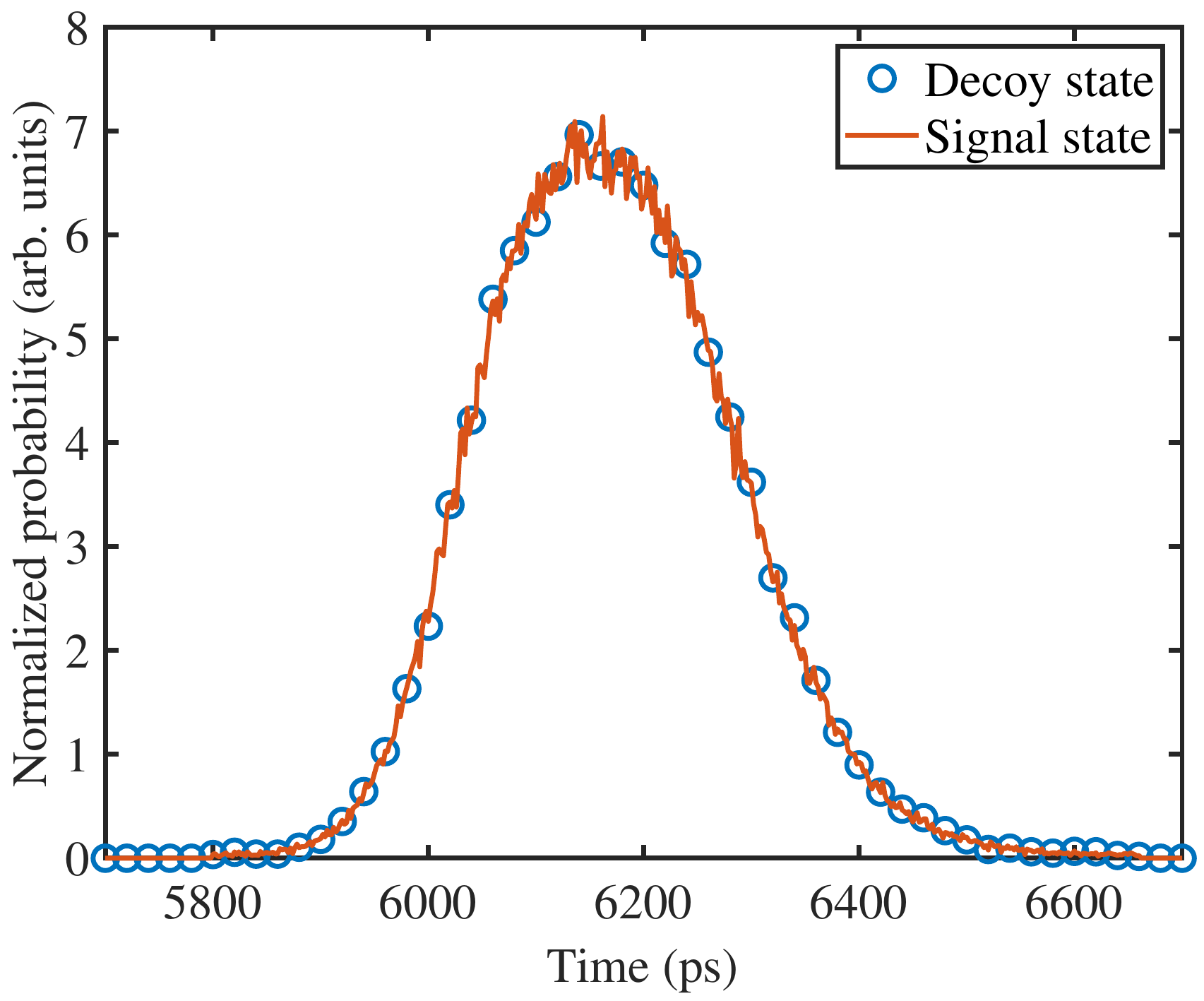}
		\caption{ Normalized intensity distribution of the signal state and the decoy state in the time domain.  The red line and blue circle represent the signal state and decoy state, respectively\blk. For comparison purposes, the signal state and decoy state were normalized to have the same area. The signal-to-decoy intensity ratio is 4:1.}\label{fig3}
	\end{figure}

	In our setup,  the signal and decoy states are modulated by Sagnac IM\blk, as shown in Fig.~\ref{fig1}. The splitter ratio of BS$_1$ was set to $75:25$. By applying the phase 0 or $\pi$ to the  $\theta$-M$_1$\blk, a signal state and decoy state with an intensity ratio of 4:1 were generated. To quantify the parameters of the distinguishable decoy states of the intensity modulation module shown in Fig.~\ref{fig1}, we directly injected the generated states into a free-running superconducting nanowire single-photon detector and measured the probability distributions of the emitted signal and decoy states at a fixed polarization state $|H\rangle$. The measurement results are shown in Fig.~\ref{fig3}. The temporal width of the laser is $400$ ps, and the distributions of the two states almost overlap. These results are expected because the  Sagnac IM \blk is an external modulation. A similar observation was made in~\cite{2018Huang-Hacking}.
	
	\begin{table*}[]
		\caption{ Probability distribution $\rho_{\mu}$ and $\rho_{\nu}$ in the time domain. We provide the normalized density matrix of the probability distribution of the pulses according to a width of 40 ps.}
		\resizebox{\linewidth}{!}{
			\begin{tabular}{cccccccccccc} 
				\hline\hline
				~~~~~~~t (ps)~~~~~~~ & ~~5800~~ & ~~5840~~ & ~~5880~~ & ~~5920~~ & ~~5960~~ & ~~6000~~ & ~~6040~~ & ~~6080~~ & ~~6120~~ & ~~6160~~ & ~~6200~~ 
				\\ \hline
				$\rho_{\mu}(t)~(10^{-4})$    & 7.732 & 8.571 & 22.71 & 72.28 & 210.7 & 489.8 & 855.3  & 1165 & 1322 & 1348 & 1283  
				\\
				$\rho_{\nu}(t)~(10^{-4})$    & 7.565 & 8.406 & 22.43 & 71.05 & 211.1 & 486.8 & 851.4  & 1166 & 1322 & 1343 & 1286 
				\\ \hline\hline
				\\ \hline\hline
				~~~t (ps)~~~ & ~6240~ & ~6280~ & ~6320~ & ~6360~ & ~6400~ & ~6440~ & ~6480~ & ~6520~ & ~6560~ & ~6600~ & ~6640~ 
				\\ \hline
				$\rho_{\mu}(t)~(10^{-4})$    & 1108 & 834.4 & 553.4 & 332.0 & 184.2 & 95.70 & 49.88 & 25.58 & 14.31 & 9.101 & 7.466  
				\\
				$\rho_{\nu}(t)~(10^{-4})$    & 1112 & 833.4 & 558.2 & 334.7 & 185.1 & 95.43 & 50.04 & 24.64 & 13.85 & 9.810 & 7.432
				\\ \hline\hline
			\end{tabular}
		}\centering \label{table_Dmunu}
	\end{table*}

	According to the method described in~\cite{2021Sun}, we use $D_{\mu \nu}=\frac{1}{2} t r\left|\rho_{\mu}(t)-\rho_{\nu}(t)\right|$ to represent the difference between the signal and decoy states, where $\rho_{\mu}(t)$ and $\rho_{\nu}(t)$ are the quantum states in which Eve is used to distinguish the signal state and decoy state for each pulse in the time domain. Based on the probability distribution trace shown in Fig.~\ref{fig2}, we can obtain the matrices of the signal and decoy states. The results of $\rho_{\mu}(t)$ and $\rho_{\nu}(t)$ are listed in Tab.~\ref{table_Dmunu}. By inputting the results in Tab.~\ref{table_Dmunu} into $D_{\mu \nu}=\frac{1}{2} t r\left|\rho_{\mu}(t)-\rho_{\nu}(t)\right|$, we obtain $D_{\mu \nu}=1.6 \times 10^{-3}$. 
	\color{black}

	\subsection{Trojan Horse}\label{Trojan-horse}
	
	\begin{figure*}
		\centering
		\resizebox{\linewidth}{!}{
			\includegraphics[width=0.9\textwidth]{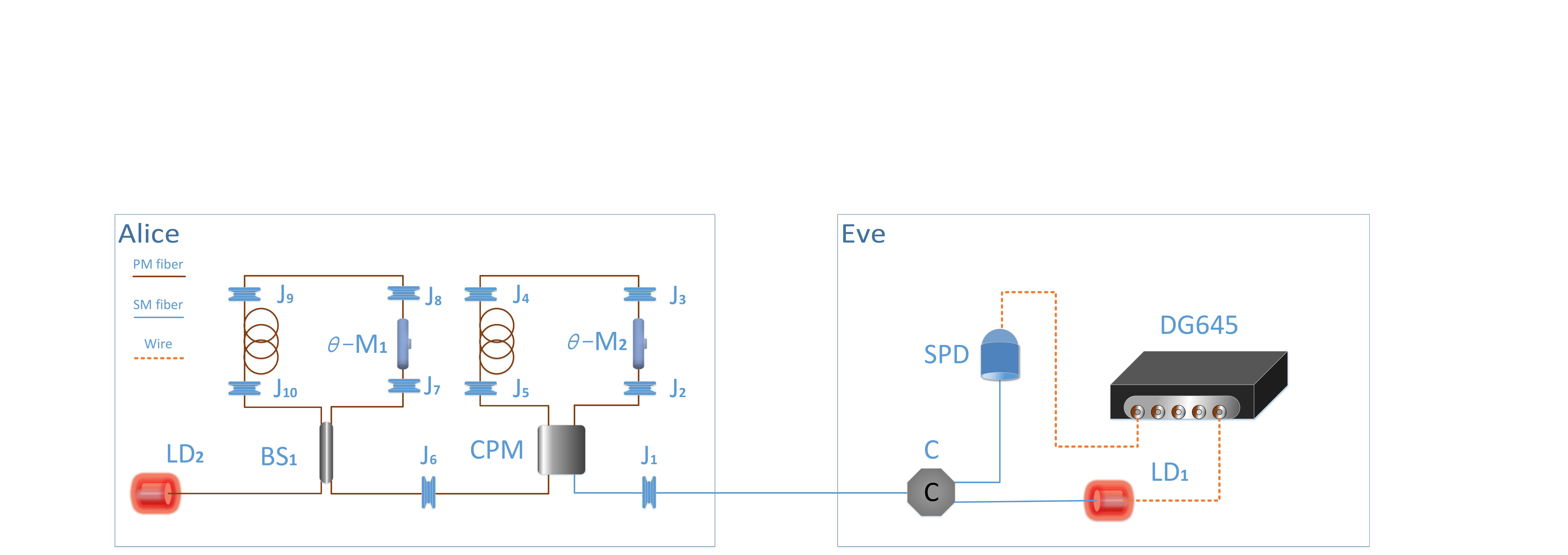}
		}
		\caption{Setup of Eve's Trojan-house attack. J$_i$: fiber flange; LD$_1$: Eve's laser; LD$_2$: Alice's laser; $\theta$-M$_i$: phase modulator; BS$_1$: beam splitter of 75:25; CPM: customized polarization module; SPD: single-photon avalanche detector; C: circulator; DG645: digital delay generator.}
		\label{fig4}
	\end{figure*}

\begin{figure}
		\centering
		\includegraphics[width=\linewidth]{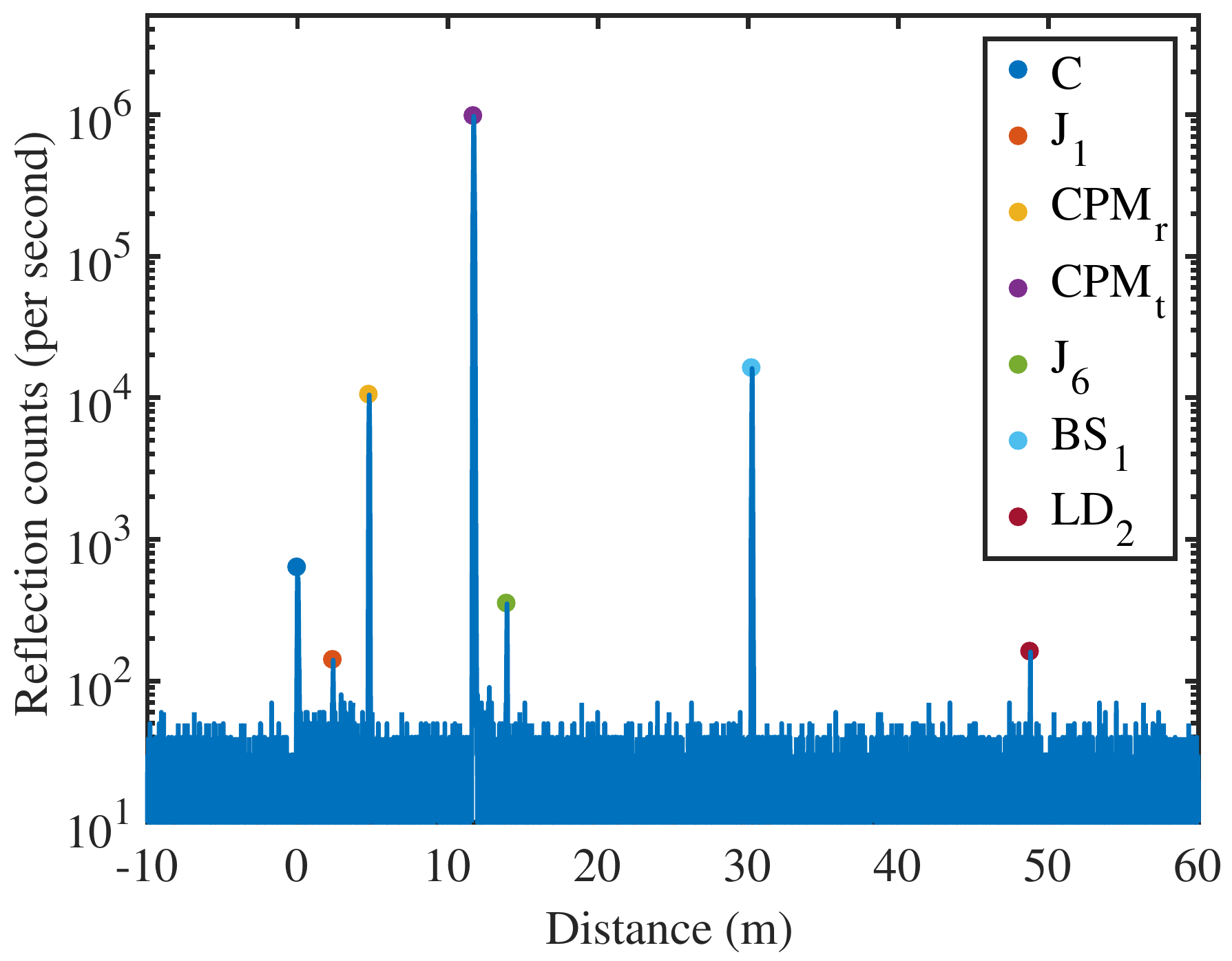}
		\caption{Leakage peaks of the transmitter unit. The distance was measured from the reflection peak of circulator C. Note that the peaks of C, J$_1$, CPM$_{\text{r}}$, J$_6$, and LD$_2$ are reflections from the end face of the device. However, the peaks of CPM$_\text{t}$ and BS$_1$ are the transmitted light due to the Trojan light entering the CPM or BS$_1$ after being transmitted in the Sagnac loop.}
		\label{fig5}
	\end{figure}

	To obtain the required parameters for characterizing the risk of Trojan-horse attacks in our system, we use a homemade single-photon optical time-domain reflectometry to measure the reflectivity of the transmitter. The measurement setup is illustrated in Fig.~\ref{fig4}, and the resulting traces are shown in Fig.~\ref{fig5}. 
	
	The schematic shown in Fig.~\ref{fig4} is as follows. A $1$ MHz pulsed laser (LD$_1$) with a center wavelength of $1550$ nm is connected to Alice through a circulator (C). Using a digital delay generator (DG645) to uniformly change the delay of a gated single-photon detector (SPD), we can obtain the resulting traces of reflected light. The measurement results are shown in Fig.~\ref{fig5}. Moreover, the time of the accumulated data for each step was 1 s.
	
	The incident light intensity of the laser was -50.251 dBm, with a corresponding number of single photons of $\mu_{Eve} = 7.360 \times 10^{10}/s$. According to Fig.~\ref{fig5}, the number of leaked $n_{leak} = 9.725 \times 10^5 /s$. Here, we include all reflection peaks that cause information leakage, that is, CPM$_t$, J$_6$, BS$_1$, LD$_2$. Subsequently, we calculate the total reflectivity $R_a$ of Alice as follows: 
	\begin{equation}
		R_a=\frac{n_{leak}}{\eta_{Eve} \times \mu_{Eve}}=1.8\times10^{-4},
	\end{equation}
	where $\eta_{Eve} = 7.34 \%$ denotes the SPD detection efficiency. Moreover, we can estimate the worst case of leaked information due to the Trojan horse attack in our setup. Suppose that Eve could attack with a number of photons up to $\mu_{Eve}^{max}=10^{20}/s$~\cite{2021PhRvP..15f4038T}, which is the maximum light intensity an optical device can withstand. Considering an optical isolator with an isolation value of 180.3dB in the transmitter, the Trojan-horse parameter $\mu_{out}$ is as follows:
	\begin{equation}
		\mu_{out}=\frac{R_a \times \mu_{Eve}^{max} \times 10^{-18.03}}{f_r}=3.6\times10^{-10},
	\end{equation}
	$f_r$ = 50 MHz is the clock frequency rate of the setup, as shown in Fig.~\ref{fig1}.
	\color{black}
	
	In the experiment, we only considered the side channel leakage caused by the Trojan horse. Thus, we can express $f$ as
	\begin{equation} \label{eqaf}
		f = \sqrt{1-\frac{\left(1- f_{t h}\right)}{\eta}}
	\end{equation}
	where $\eta = \eta_{c} \times \eta_{B o b} \times \eta_{d}$, and $f_{th}$ is the Trojan horse fidelity matrix, which can be expressed as
	\begin{equation} \label{eqafth}
		\begin{aligned}
			f_{t h} &=\left[\begin{array}{cccc}
				1 & e^{-4 \mu_{out }} & e^{-2 \mu_{out }} & e^{-2 \mu_{out }} \\
				e^{-4 \mu_{out }} & 1 & e^{-2 \mu_{out }} & e^{-2 \mu_{out }} \\
				e^{-2 \mu_{out }} & e^{-2 \mu_{out }} & 1 & e^{-4 \mu_{out }} \\
				e^{-2 \mu_{out }} & e^{-2 \mu_{out }} & e^{-4 \mu_{out }} & 1 
			\end{array}\right] ,
		\end{aligned}
	\end{equation}
	Here, we consider the upper bound of $\mu_{out}$, and thus, we have
	\begin{equation} \label{eqafth_date}
		\begin{aligned}
			&f_{t h}=\\
			&\left[
			\begin{array}{cccc}
				1 & e^{-1.4 \times 10^{-9}} & e^{-7.2 \times 10^{-10}} & e^{-7.2 \times 10^{-10}} \\
				e^{-1.4 \times 10^{-9}} & 1 & e^{-7.2 \times 10^{-10}} & e^{-7.2 \times 10^{-10}} \\
				e^{-7.2 \times 10^{-10}} & e^{-7.2 \times 10^{-10}} & 1 & e^{-1.4 \times 10^{-9}} \\
				e^{-7.2 \times 10^{-10}} & e^{-7.2 \times 10^{-10}} & e^{-1.4 \times 10^{-9}} & 1 
			\end{array}
			\right].
		\end{aligned}
	\end{equation}
	Then, we can give the side channel matrix $f_{\beta_{i}}$ used in Eq.~\eqref{eq3} as follows:
	\begin{equation} \label{eqa16}
		\begin{aligned}
			f_{z_{0}}&=diag \left[ f_{(1, 1)} \quad f_{(1, 2)} \quad f_{(1, 3)} \quad f_{(1, 4)} \right]
			\\
			f_{z_{1}}&=diag \left[ f_{(2, 1)} \quad f_{(2, 2)} \quad f_{(2, 3)} \quad f_{(2, 4)} \right]
			\\
			f_{x_{0}}&=diag \left[ f_{(3, 1)} \quad f_{(3, 2)} \quad f_{(3, 3)} \quad f_{(3, 4)} \right]
			\\
			f_{x_{1}}&=diag \left[ f_{(4, 1)} \quad f_{(4, 2)} \quad f_{(4, 3)} \quad f_{(4, 4)} \right] ,
		\end{aligned}
	\end{equation}
	where $f_{(i,j)}$ is the element in the i-th row and j-th column of the matrix $f$.
	
	According to $f$, $f_{\beta_{i}}$, and $Z_{ij}$, we can obtain the equation for $Z_{ij}^p$ as follows:
	\begin{equation} \label{eqa5}
		\begin{aligned}
			Z_{ij}^{p} &=
			\left(V_{ij}^{0}\right)^{+} \cdot f \cdot V_{ij}^{0}+\left(V_{ij}^{1}\right)^{+} \cdot f \cdot V_{ij}^{1} \\
			&+\left(V_{ij}^{0}\right)^{+} \cdot f \cdot V_{ij}^{1}+\left(V_{ij}^{1}\right)^{+} \cdot f \cdot V_{ij}^{0} ,
		\end{aligned}
	\end{equation}
	where $i,j=0,1$, $V_{ij}^{0}=\Omega_{ij}^{0} \otimes f_{z_{0}}$, and $V_{ij}^{1}=\Omega_{ij}^{1} \otimes f_{z_{1}}$. $\Omega_{ij}$ can be expressed as follows:
	\begin{equation} \label{eqa6}
		\begin{aligned}
			&\Omega_{00}^{0}=Z_{00}+Z_{01}, &\Omega_{00}^{1}=Z_{10}+Z_{11}\\
			&\Omega_{01}^{0}=Z_{00}-Z_{01},	&\Omega_{01}^{1}=Z_{10}-Z_{11}\\
			&\Omega_{10}^{0}=Z_{00}+Z_{01},	&\Omega_{10}^{1}=-Z_{10}-Z_{11}\\
			&\Omega_{11}^{0}=Z_{00}-Z_{01},	&\Omega_{11}^{1}=-Z_{10}+Z_{11} .
		\end{aligned}
	\end{equation}

	\subsection{Detection mismatch}\label{Detection mismatch}

	Here, we consider Eve's time-shift attack~\cite{PhysRevA.78.042333} on Bob's detectors, which implies that Eve can randomly shift the arrival time of each signal to either $t_1$ or $t_2$. Thus, Bob's measurement result is biased toward 0 or 1 depending on the arrival time $t_1$ or $t_2$.

\begin{figure}
		\centering
		\subfigure[]{
			\includegraphics[width=\linewidth]{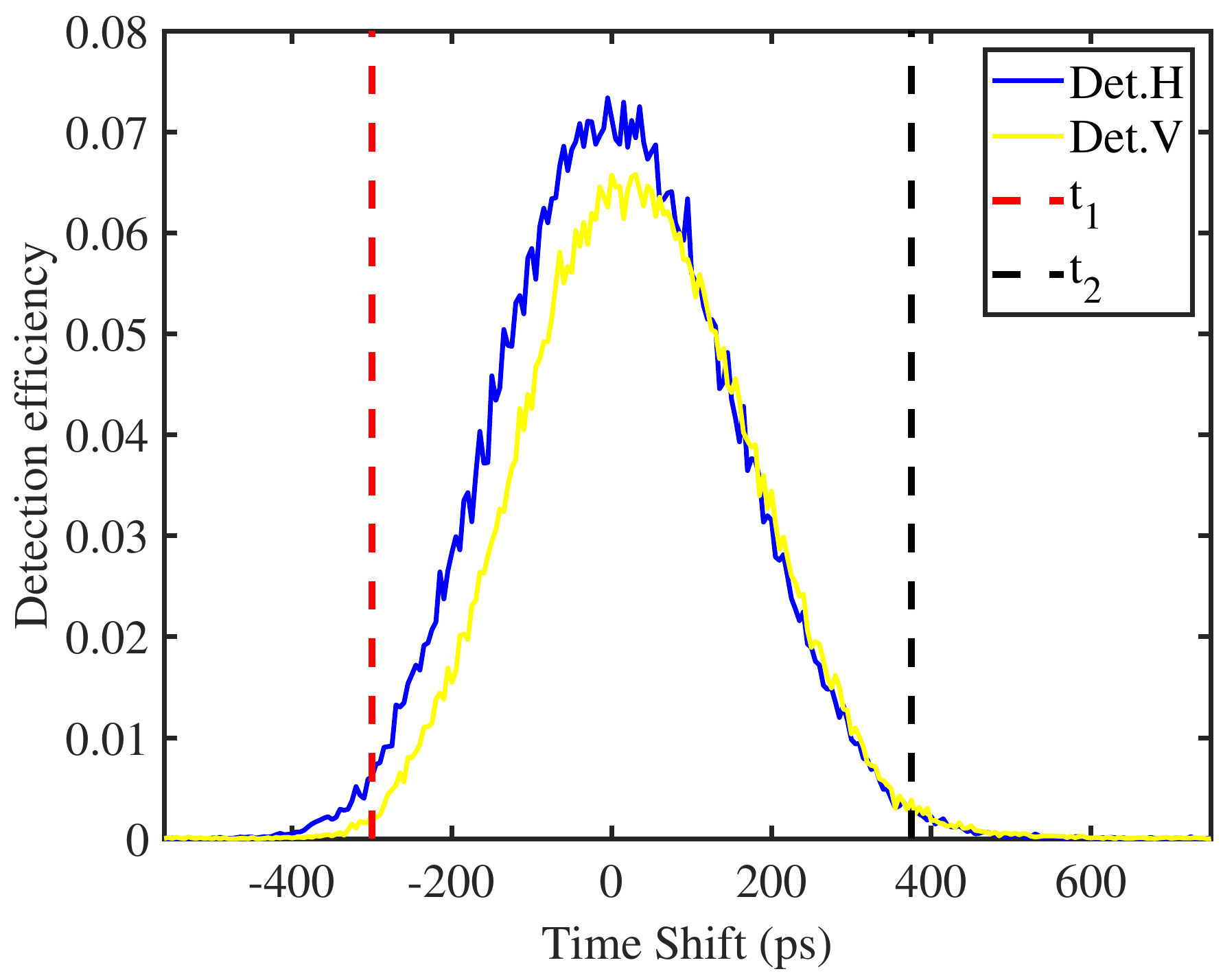} 
		}
		\subfigure[]{
			\includegraphics[width=\linewidth]{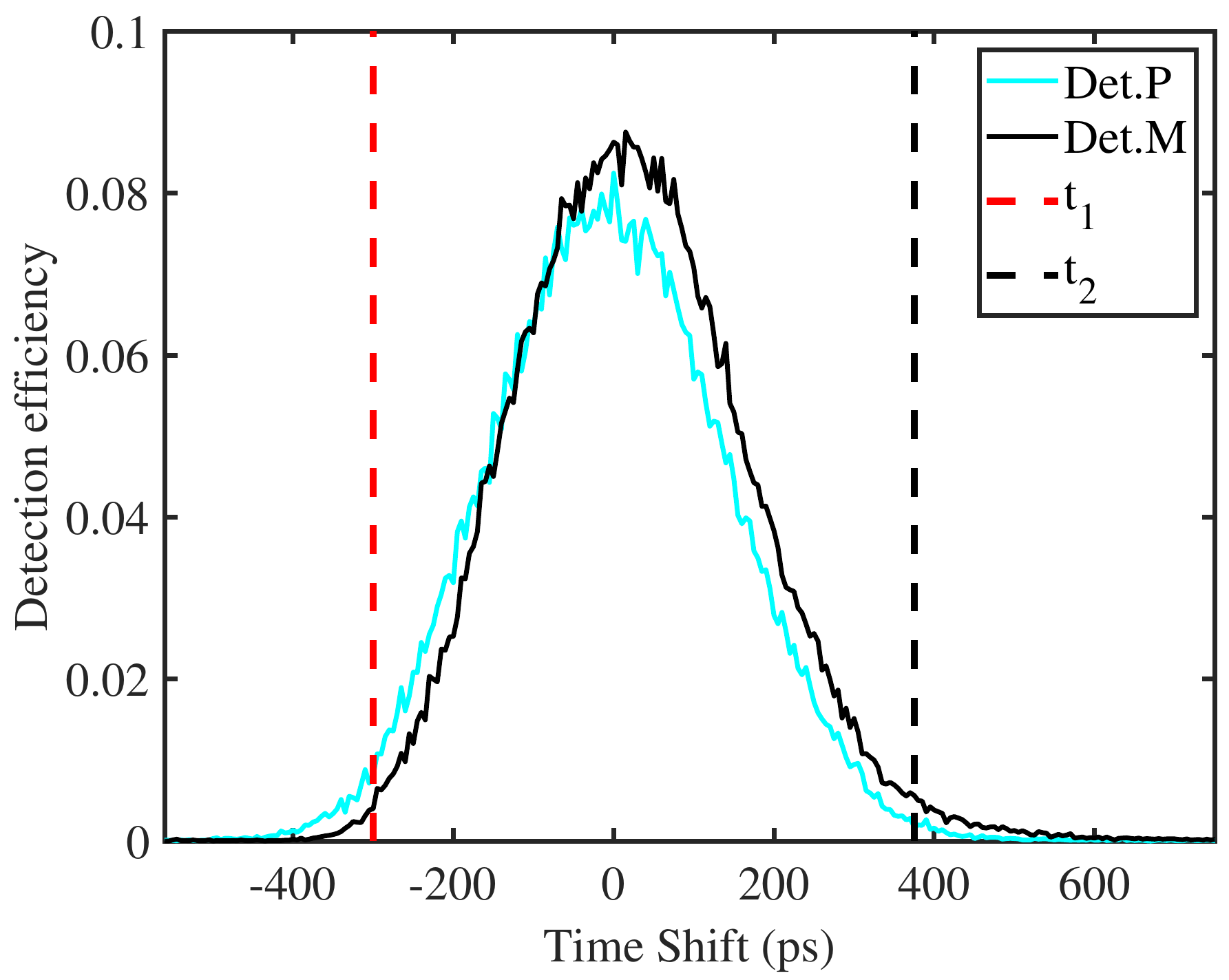} 
		}
		\caption{ Efficiencies of four detectors versus time shifts. (a)  Detection efficiency of the $Z$-basis, blue and yellow lines represent the detector H and detector V. (b) Detection efficiency on the $X$-basis, cyan and black lines represent the detector P and detector M. $t_1$ (dashed red line) =-300 ps and $t_2$ (dashed black line) =375 ps are the optimal shift times for the attack\blk.}
		\label{fig7}
	\end{figure}

	To obtain the flaw parameters $F_j^+ F_j$ and $C$ of our system under a time-shift attack, we measured the detection efficiency of the four detectors under the same light pulse. This approach ensured the consistency of the pulse arrival time. The light intensity used in the experiment was -104.176dBm, and the gating of SPD was 1ns. By using DG645 to change the delay of SPD at 5 ps steps uniformly, we can obtain the detection efficiency curve of the four detectors. The results are shown in Fig.~\ref{fig7}.

	The measurements of the proposed system are performed using four detectors, limiting Eve's time-shift attack to some extent. In this case, Eve cannot determine Bob's basis of measurement; therefore, Eve cannot simultaneously select the optimal attack for each basis simultaneously. Eve could select large shifts in $Z$-basis and $X$-basis by a simple consideration, which would provide substantial intrinsic detection efficiency mismatches. Note that the operation of Bob to select the measurement basis is independent of Eve's time-shift attack; therefore, Eve should have the same arrival time for different bases in Fig.~\ref{fig7}. In this case, we provide the optimal pulse arrival time of Eve by considering the detection mismatch between the $Z$-basis and the $X$-basis. The detection efficiencies of the four detectors at  $t_1=-300$ and $t_2=375$ \blk are listed in Tab.~\ref{table4}.
	
	We define H and P polarization states as bit $``0"$ and V and M polarization states as bit $``1"$. Moreover, the probability of choosing the $Z$-basis is $90\%$. Based on the values listed in Tab.~\ref{table4}, we can obtain the detection efficiency matrix $F_j$ used in Eq.~\eqref{eq3},

	\begin{table*}[]
		\centering 
		\caption{  Experimental raw counts. $n_{\beta, k}$ and $m_{\beta, k}$ are the detection and error detection events on the $\beta$-basis with a specific intensity $k\in\{\mu,\nu\}$. $n_{\beta_i, k}$ and $m_{\beta_i, k}$ are the parts of $n_{\beta, k}$ and $m_{\beta, k}$ that Alice sends encoded as a bit $``i"$ pulse, $i\in{0,1}$. $N_{\beta,\beta'}$ is the total number of pulses that Alice sends the quantum state in $\beta$-basis and Bob successfully detects in the $\beta'$-basis. $N_{\beta_i,\beta'}$ represents the counts when the total number of pulses that Alice sends the quantum state in the $\beta$-basis with bit $``i"$ and Bob detects in the $\beta'$-basis, ${\beta,\beta'}\in\{z,x\}$.}
		\resizebox{\linewidth}{!}{
			\begin{tabular}{cccccccccccccc} 
				\hline\hline
				$L(km)$ & $n_{z_0,\mu}$ & $m_{z_0,\mu}$ & $n_{z_1,\mu}$ & $m_{z_1,\mu}$ & $n_{x_0,\mu}$ & $m_{x_0,\mu}$ & $n_{x_1,\mu}$ & $m_{x_1,\mu}$ & $n_{z,\mu}$ & $m_{z,\mu}$ & $N_{zz} (10^9)$ & $N_{z_0z} (10^9)$ & $N_{z_1z} (10^9)$ 
				\\ \hline
				25    & 55776166 & 549127 & 59132433 & 670342 & 635779 & 7764 & 641408 & 5836 & 114908599 & 1219469 & 8.1529 & 4.0764 & 4.0764
				\\
				50    & 13572261 & 147390 & 15000021 & 133335  & 137787 & 1708 & 169011 & 2109 & 28572282 & 280725 & 8.1296 & 4.0648 & 4.0648
				\\
				75    & 4576860  & 54744  & 3828257  & 50017  & 72260  & 1310 & 76612  & 1600 & 8405117 & 104761 & 8.1390 & 4.0695 & 4.0695
				\\ \hline\hline
				\\ \hline\hline
				$L(km)$ & $n_{z_0,\nu}$ & $m_{z_0,\nu}$  & $n_{z_1,\nu}$ & $m_{z_1,\nu}$ & $n_{x_0,\nu}$ & $m_{x_0,\nu}$ & $n_{x_1,\nu}$ & $m_{x_1,\nu}$ & $n_{z,\nu}$ & $m_{z,\nu}$ &$N_{xx} (10^7)$ & $N_{x_0z} (10^7)$ & $N_{x_1x} (10^7)$ 
				\\ \hline
				25    & 3192913 & 84869 & 2166169 & 29448 & 33203 & 825 & 34917 & 514  & 5359082 & 114317 & 10.065 &5.0326 & 5.0226 
				\\
				50    & 1185426 & 27400 & 961101 & 9283 & 18647 & 549  & 11265 & 177 & 2146527 & 36683 & 10.036 & 5.0182 & 5.0182 
				\\
				75    & 1078178 & 17235 & 965843  & 17583 & 13072 & 458  & 12381 & 387 & 2044021 & 34818 & 10.048 & 5.0240 & 5.0240 
				\\ \hline\hline
			\end{tabular}
		}
		\label{table_rawdate}
	\end{table*}

	\begin{table}[]
		\centering
		\caption{ Detection efficiency in $t_1$ and $t_2$. $\eta_{H}$, $\eta_{V}$, $\eta_{P}$, and $\eta_{M}$ are the detection efficiencies of detectors H, V, P, and M, respectively. Moreover, $t_1$ and $t_2$ are the pulse arrival times. }
		\begin{tabular}{cccccc||cccccc}
			\hline \hline
			&~~&   $t_1$   &~~&   $t_2$   &~~&&               &~~&   $t_1$   &~~&   $t_2$     \\ \hline
			$\eta_{H}(t)$  &  &  0.55\%  & &  0.33\% &  &&  $\eta_{P}(t)$   & & 0.71\% & &  0.18\%   \\
			$\eta_{V}(t)$  &  &  0.31\%  & &  0.38\%  &  &&  $\eta_{M}(t)$   & & 0.51\% & &  0.56\%   \\ \hline \hline  
		\end{tabular}
		\label{table4}
	\end{table}
	
	\begin{equation} \label{eqa_Fj}
		F_{j}=0.9 \times F_{z_j}+0.1 \times F_{x_j},
	\end{equation}
	here $F_{z_j}$ and $F_{x_j}$ are the normalized matrices of detection efficiency of bit $``j"$ in the $Z$-basis and $X$-basis respectively, which can be expressed as follows:
	\begin{equation} \label{eqa_Fzxj}
		\begin{aligned}
			& F_{z_0}=\left[\begin{array}{ccc} 1 &~& 0 \\ 0&~& \frac{\eta_{H}(t_2)}{\eta_{V}(t_2)} \end{array}\right]
			& F_{z_1}=\left[\begin{array}{ccc} \frac{\eta_{V}(t_1)}{\eta_{H}(t_1)} &~& 0 \\ 0&~& 1 \end{array}\right]~
			\\
			& F_{x_0}=\left[\begin{array}{ccc} 1 &~& 0 \\ 0&~& \frac{\eta_{P}(t_2)}{\eta_{M}(t_2)} \end{array}\right]
			& F_{x_1}=\left[\begin{array}{ccc} \frac{\eta_{M}(t_1)}{\eta_{P}(t_1)} &~& 0 \\ 0&~& 1 \end{array}\right] .
		\end{aligned}
	\end{equation}
	By substituting the detection efficiencies listed in Tab.~\ref{table4} into Eq.~\eqref{eqa_Fj} and Eq.~\eqref{eqa_Fzxj}, we can obtain $F_0$ and $F_1$ such that 
	\begin{equation}
		F_{0}=\left[\begin{array}{ccc} 1 &~& 0 \\ 0&~& 0.8137 \end{array}\right]~~
		F_{1}=\left[\begin{array}{ccc} 0.5791 &~& 0 \\ 0&~& 1 \end{array}\right].
	\end{equation}
	\color{black}

	Subsequently, we can use $F_0$ and $F_1$ to construct $C$ in Eq.~\eqref{eq3}. First, we build an auxiliary matrix $CF$ as follows: 
	\begin{equation} \label{eqa17}
		CF=F_{0}\left(F_{1}{ }^{+} F_{1}\right)^{-1} F_{0}{ }^{+}.
	\end{equation}
	All eigenvectors and eigenvalues of $CF$ are composed of $V$ and $D$,
	
	\begin{equation} \label{eqa18}
		\begin{aligned}
			D& =diag \left[\begin{array}{cccc}
				\lambda_{1} & \lambda_{2} & \cdots & \lambda_{n} \\
			\end{array}\right] \\
			&V =\left[\begin{array}{llll}
				\varphi_{1} & \varphi_{2} & \cdots & \varphi_{n}
			\end{array}\right] ,
		\end{aligned}
	\end{equation}
	where $\lambda_{n}$ is the $n$-th eigenvalue of matrix $CF$, and $\varphi_{n}$ is the $n$-th eigenvector. Then, $C$ can be expressed as
	follows:
	\begin{equation} \label{eqa19}
		C=V^{+} F_{0} \sqrt{\min \left[\frac{1}{D}, diag(n)\right]} ,
	\end{equation}
	where $diag(n)$ is a unitary diagonal matrix equal to dimension $D$.

	\section{Experimental data}\label{AppendixB}
	
	Detailed experimental data are listed in Tab.~\ref{table_rawdate}.


	%

\end{document}